\documentclass{IEEEtran}

\usepackage{times}

\usepackage[T1]{fontenc}
\usepackage{amsmath,amsfonts,amssymb,dsfont,subfigure,amssymb,vaucanson-g,graphicx,color}%,cite}
\usepackage{authblk}
\usepackage{vaucanson-g}
\usepackage{tikz}
\usetikzlibrary{arrows}

\newcommand{\rea}{\mathbb{R}}

\newtheorem{theorem}{\textbf{Theorem}}

\newtheorem{lemma}[theorem]{\textbf{Lemma}}

\author{Mehran Zareh$^*$, Dimos V. Dimarogonas$^{**}$, Mauro Franceschelli$^*$,\\ Karl Henrik Johansson$^{**}$, Carla Seatzu$^*$
\\~\\
$^{*}$ Department of Electrical and Electronic Engineering, University of Cagliari, Italy \\
\{mehran.zareh,mauro.franceschelli,seatzu\}@diee.unica.it \\~\\
$^{**}$ School of Electrical Engineering, Royal Institute of Technology, Stockholm, Sweden \\ \{dimos,kallej\}@kth.se}

\title{\LARGE \bf
Consensus in multi-agent systems with second-order dynamics \\ and non-periodic sampled-data exchange}
\begin{document}
\maketitle
%\setcounter{page}

%\fontencoding{U}\fontfamily{psy}\selectfont ABCDEFGH

\section*{Abstract} In this paper consensus in second-order multi-agent
systems with a non-periodic sampled-data exchange among agents is investigated. The sampling is random with bounded inter-sampling intervals. It is assumed that each agent has exact knowledge of its own state at all times. The considered local interaction rule is PD-type. The characterization of the convergence properties exploits a Lyapunov-Krasovskii functional method, sufficient conditions for stability of the consensus protocol to a time-invariant value are derived. Numerical simulations are presented to corroborate the theoretical results.

\section{Introduction}
In the past years, a significant attention has been devoted to the consensus problem in multi-agent systems
 (MAS) \cite{qin2011second,ren2005survey,yu2010some,Zareh_consensus2} by the scientific community.
Potential applications of consensus algorithms are found in sensor networks \cite{yu2009distributed,olfati2005consensus}, automated highway systems \cite{ren2005survey}, mobile robotics \cite{khoo2009robust} and satellite alignment \cite{ren2007distributed}. In particular, the coordination problem of mobile robots finds several applications in the manufacturing industry in the context of automated material handling.

The consensus problem in the context of mobile robots consists in the design of local state update rules which allow the network of robots to rendezvous at some point in space or follow a leading robot exploiting only measurements of speeds and relative positions between neighboring robots. Robots are hereafter referred to as agents.

%In the consensus literature usually the dynamics of the agents are considered to be of first order .

%{\red ... rewrite ... Due to fast sensation and computation nowadays, we can suppose the agents can have their own local data almost with no delay. Therefore it is a reasonable estimation to consider that the local states of the agents are continuous. It can be of interest if the communication among the agents occurs as less as possible. Communication costs and if it could be reduced somehow then one can save more energy. But what would happen if an agent receives information in a non continuous manner. Obviously it may cause if we suppose that the receiving information is held till the next communication occurs then the receiving agent will look at the past data of its neighbors and this may cause the system not to reach consensus. ...}

In this paper, we consider the case in which each agent has a perfect knowledge of its own state with almost no delay, i.e., it knows its own speed and position. Information exchanges between neighboring agents happens at discrete time intervals which are possibly non-periodic but strictly positive and bounded.   %Information exchange
%among agents also occur immediately. Here we suppose that once an information
%on a given agent is received by another agent, then it is being hold as the current state of the neighbors until new information arrives.
The network dynamics can thus be modeled as a \emph{sampled-data system} (SDS), a class of systems extensively investigated in the literature.

For interesting contributions in this area we point the reader to \cite{ackermann1985sampled,fridman2010refined,zutshi2012timed} and the references therein. We also mention the work by Fridman {\em et al.} \cite{fridman2004robust} who exploited an approach for time-delay systems and obtained the sufficient stability conditions based on the Lyapunov-Krasovskii functional method. In Seuret \cite{seuret2012novel} and Fridman \cite{fridman2010refined} improved methods with better upper bounds to the maximum allowed delay were proposed. Shen {\em et al.} \cite{shen2012sampled} studied the sampled-data synchronization control problem for dynamical networks. Qin {\em et al.} \cite{qin2010sampled} and Ren and Cao \cite{ren2008convergence} studied the consensus problem for networks of double integrators with a constant sampling period. In the latter two papers, even though the authors use the sampled-data notation to introduce their novelty, they suppose that the communication and the local sensing occur simultaneously and this simplifies the problem into a discrete state consensus problem. Xiao and Chen \cite{xiao2012sampled} and Yu {\em et al.} \cite{yu2011second} studied second-order consensus in multi-agent dynamical systems with sampled \emph{position} data.

In this paper we propose a PD-like consensus algorithm with non-periodic sampled-data exchange among agents with bounded and strictly positive inter-sampling intervals. A characterization of the convergence properties exploiting a Lyapunov-Krasovskii functional method is provided and sufficient conditions for exponential stability of the consensus protocol to a time-invariant value are derived. Numerical simulations are presented to corroborate the theoretical results.

The paper is organized as follows. In Section~\ref{section:Prob_for} some notation and preliminaries are introduced. In Section~\ref{ProblemState} the consensus problem for second order multi-agent systems with non-periodic sampled-data exchange is formalized. In Section~\ref{main_res_1} the convergence properties of the proposed consensus protocol are characterized. In Section~\ref{sim_res} simulation results are presented to corroborate the theoretical analysis. In Section~\ref{conclusions} concluding remarks and directions for future research are discussed.

\section{Notation and Preliminaries}\label{section:Prob_for}

In this section we recall some basic notions on graph theory and introduce the notation used in the paper.

The topology of bidirectional communication channels
among the agents is represented by an undirected graph $\mathcal{G}=(\mathcal{V}, \mathcal{E})$ where $\mathcal{V}=\{1,\ldots,n\}$ is the set of nodes (agents) and $\mathcal{E}\subseteq\{\mathcal{V}\times \mathcal{V}\}$ is the set of edges. An edge $(i, j) \in  \mathcal{E}$ exists if there is a
communication channel between agent $i$ and $j$. Self loops
$(i, i)$ are not considered. The set of neighbors of agent $i$
is denoted by $\mathcal{N}_i  = \{j \ : \ (j, i) \in  \mathcal{E}; j = 1, \ldots, n\}$. Let
$\delta_i = |\mathcal{N}_i|$ be the degree of agent $i$ which represents the total
number of its neighbors.

The topology of graph $\mathcal{G}$ is encoded by the so-called {\em adjacency matrix}, an $n \times n$ matrix $A_d$ whose $(i,j)$-th entry is equal to $1$ if $(i,j)\in \mathcal{E}$, $0$ otherwise. Obviously in an undirected graph matrix $A_d$ is symmetric.

We denote $\Delta=\emph{diag}(\delta_{1}, \ldots,\delta_{n})$ the diagonal matrix whose non null entries are the degrees of the nodes. Moreover, matrix $W_d=\Delta^{-1} A_d$ is the {\em weighted adjacency matrix} associated with $\mathcal{G}$.
The following result holds.

\begin{lemma}\label{lemma_connected}
If a graph $\mathcal{G}$ is connected then the eigenvalues of the weighted adjacency matrix $W_d$, namely $\lambda_i, \ \, i=1,\ldots,n$, are all located in the interval $[-1, \ 1]$, and $\lambda_1=1$ is always a simple eigenvalue of $W_d$.
\end{lemma}

{\em Proof:} Using Gershgorin theorem since all the diagonal elements of $W_d$ are zero and each row sums up to $1$, it immediately follows that $\lambda_i \in[-1, \ 1]$. Now, let $L=\Delta-A_d$ be the Laplacian matrix associated with the considered graph. If such a graph is connected, then the origin is a simple eigenvalue of $L$ which implies that  it is a simple eigenvalue also for $-\Delta^{-1}L=\Delta^{-1}A_d-I=W_d-I$. Consequently, if the graph is connected, $\lambda_1=1$ is a simple eigenvalue of the weighted adjacency matrix.

\hfill $\square$

%{\red ... is it already well know? shall we keep the proof? ....}

Finally, in the rest of this paper we denote with $*$ the symmetric elements of symmetric matrices.

\section{Problem Statement}\label{ProblemState}

Consider a second-order multi-agent system with an undirected communication topology. Consider the PD-type consensus protocol inspired by  \cite{cepeda2011exact} and \cite{Zareh_consensus}:
\begin{equation}\label{maineq_not_sampled}
\left\{
\begin{array}{lll}
 \dot{x}_i(t) & = & v_i(t),
\\
\dot{v}_i(t)& = & \dfrac{k_p}{\delta_i}\sum_{j \in \mathcal{N}_i}x_j(t)+\dfrac{k_d}{\delta_i}\sum_{j \in \mathcal{N}_i}v_j(t) \\
 & & -k_p x_i(t)-k_d v_i(t),
\end{array}
\right.
\end{equation}
where $i=1, \ldots, n$, $n$ denotes the number of agents, $x_i(t)$ and $v_i(t)$ are the position and the velocity of agent $i$, and $\delta_i$ indicates its degree.

We suppose that the local information, i.e., the information that each agent receives from its own sensors, is measured instantaneously. This obviously makes sense when the sensor dynamics are fast enough. %The neighbors' data instead is captured in a arbitrary sampled process while each two successive samplings do not occur later than a known constant, $\bar{\tau}$.

Moreover, we assume that the communication between the generic agent $i$ and its set of neighbors $\mathcal{N}_i$ occurs in stochastic sampling time instants $t_k$, $k=0,1,\ldots, \infty$ that satisfy the following conditions: $$0  < t_{k+1}-t_k\leq \bar{\tau}\in \mathbb{R}^+$$ and $$\lim\limits_{k\to \infty}t_k=\infty.$$

Under the above assumptions, equation \eqref{maineq_not_sampled} can be rewritten as:
\begin{equation}\label{eq_2_c}
\left\{
\begin{array}{lll}
 \dot{x}_i(t) & = & v_i(t),
\\
\dot{v}_i(t) & =  & \dfrac{k_p}{\delta_i}\sum_{j \in \mathcal{N}_i}x_j(t_k)+\dfrac{k_d}{\delta_i}\sum_{j \in \mathcal{N}_i}v_j(t_k) \\
 & & -k_p x_i(t)-k_d v_i(t)
\end{array}
\right.
\end{equation}
or, alternatively, doing some simple manipulations, as:
\begin{equation}\label{main}
\left[\begin{array}{c}
\dot{x}(t) \\
\dot{v}(t)
\end{array} \right]=(A\otimes I_n)\left[\begin{array}{c}
{x}(t) \\
{v}(t)
\end{array}\right]+(B\otimes W_d)\left[\begin{array}{c}
{x(t_k)} \\
{v(t_k)}
\end{array}\right]
\end{equation}
where $t\in [t_k,t_{k+1})$, $x=[x_1,x_2,\ldots, x_n]$, $v=[v_1,v_2,\ldots,v_n]$, $\Delta=\emph{diag}\{\delta_1,\delta_2,\ldots, \delta_n\}$, $A_d$ is the adjacency matrix, $W_d=\Delta^{-1}A_d$ is the weighted adjacency matrix, and matrices $A $ and $B$ are equal, respectively, to:
\begin{equation}\label{def:AB}
A=\left[\begin{array}{cc}
0 &1  \\
-k_p &-k_d
\end{array} \right], \qquad B=\left[\begin{array}{cc}
0 &0  \\
k_p &k_d
\end{array} \right].
\end{equation}

A MAS with an undirected communication topology and following equation~\eqref{maineq_not_sampled}, is said to converge to a \emph{consensus state} if $$\lim\limits_{t \to \infty }|x_i(t)-x_j(t)|=0$$ and $$\lim\limits_{t \to \infty }|v_i(t)-v_j(t)|=0.$$

In this paper, given the value of the maximum admissible difference $\bar{\tau}$ between any two consecutive sampling time instants, and a communication topology with a given spectrum, we aim at finding conditions that guarantee consensus to a fixed point among agents that evolve according to equation~\eqref{main}.

We will also address the issue of evaluating an upper bound to the decay rate of convergence.

We conclude this section pointing out some differences among our problem formulation and the ones in \cite{xiao2012sampled} and \cite{yu2011second}. The most important difference is that we assume that each agent receives a message containing its neighbors' positions and velocities in a sampled-data basis. On the contrary, both in \cite{xiao2012sampled} and in \cite{yu2011second}, the agents gather the sampled positions of their neighbors and their own at the same time instants.

\section{Convergence properties}\label{main_res_1}

In the following subsection we first introduce a state variable transformation to decouple the dynamics of modes associated with the eigenvalues of the weighted adjacency matrix. Then, the stability of such modes is analyzed in detailed.

\subsection{Stability analysis}

Apply the following change of variables:
\begin{equation}
x(t)=T z(t)
\end{equation}
to eq.~\eqref{main}. Then, it holds:
\begin{equation}\label{transformed}
\begin{array}{lll}
(I_2 \otimes T)\left[\begin{array}{c}
\dot{z}(t) \\
\ddot{z}(t)
\end{array} \right] & = & (A\otimes  T)\left[\begin{array}{c}
{z}(t) \\
{\dot{z}}(t)
\end{array}\right] \\ & & +(B\otimes W_d T)\left[\begin{array}{c}
{z(t_k)} \\
{\dot{z}(t_k)}
\end{array}\right]
\end{array}
\end{equation}
and eq.~\eqref{main} can be rewritten as:
\begin{equation}\label{eq1_c}
\begin{array}{lll}
\left[\begin{array}{c}
\dot{z}(t) \\
\ddot{z}(t)
\end{array} \right] & = & (A\otimes I_n)\left[\begin{array}{c}
{z}(t) \\
{\dot{z}}(t)
\end{array}\right] \\ & & +(B\otimes T^{-1}W_d T)\left[\begin{array}{c}
{z(t_k)} \\
{\dot{z}(t_k)}
\end{array}\right].
\end{array}
\end{equation}
Since $W_d$ is a symmetrizable matrix, then it is also diagonalizable \cite{cepeda2011exact}, and the transformation matrix $T$ can be chosen such that $$\Lambda=T^{-1}W_d T=\emph{diag}(\lambda_1, \lambda_2, \ldots, \lambda_n)$$ where $$\lambda_1\geq \lambda_2\geq \ldots\geq \lambda_n$$ are the eigenvalues of the weighted adjacency matrix $W_d$.
As a result, eq.~\eqref{eq1_c} can be rewritten as:
$$\left[\begin{array}{c}
\dot{z}(t) \\
\ddot{z}(t)
\end{array} \right]=(A\otimes I_n )\left[\begin{array}{c}
{z}(t) \\
{\dot{z}}(t)
\end{array}\right]+(B\otimes \Lambda)\left[\begin{array}{c}
{z(t_k)} \\
{\dot{z}(t_k)}
\end{array}\right],$$
or alternatively, as
\begin{equation}
\left[\begin{array}{c}
\dot{z}_i(t) \\
\ddot{z}_i(t)
\end{array} \right]=A\left[\begin{array}{c}
{z}_i(t) \\
{\dot{z}}_i(t)
\end{array}\right]+\lambda_i B  \left[\begin{array}{c}
{z_i(t_k)} \\
{\dot{z}_i(t_k)}
\end{array}\right]
\end{equation}
where $i=1,\ldots,n$, and $z_i(t)$ is the $i$-th element of vector $z(t)$.

Now, if we define
\begin{equation}\label{def:mode}
y_i(t)=[z_i(t)\ \ \dot{z}_i(t)]^T
\end{equation}
the $i$-th {mode} of the system, we can say that its dynamics follows equation:
\begin{equation}\label{eq1}
\dot{y}_i(t)=Ay_i(t)+\lambda_i By_i(t_k).
\end{equation}

Moreover, assuming $\tau(t)=t-t_k$, the above equation can be rewritten as:
\begin{equation}\label{mode_dynamics}
\dot{y}_i(t)=Ay_i(t)+\lambda_i By_i(t-\tau(t)).
\end{equation}
The above SDS is a special case of a time varying delayed system where the delay $\tau(t)$ is upper bounded by $\bar{\tau}$, and its derivative is $\dot{\tau}(t)=1$, while the delay switches at times $t=t_k$, $k=0,1,\ldots, \infty$.

%The following result that will be useful in the rest of the paper can be proved.
%
%\begin{lemma}
%If a graph $\mathcal{G}$ is connected then the eigenvalues of the weighted adjacency matrix $W_d$, namely $\lambda_i, \ \, i=1,\ldots,n$, are all located in the interval $[-1\ \ 1]$, and $\lambda_1=1$ is always a simple eigenvalue of $W_d$.
%\end{lemma}
%\begin{proof}
%Using Gershgorin theorem since all the diagonal elements of $W_d$ are zero and each row sums up to $1$, it immediately follows that $\lambda_i \in[-1\ \ 1]$.
%
%{\blue Now, let $L=\Delta-A_d$ be the Laplacian matrix associated with the considered graph. If such a graph is connected, then $\bar \lambda =0$ is a simple eigenvalue of $L$. This implies that $\bar \lambda=0$ is also a simple eigenvalue for $-\Delta^{-1}L=\Delta^{-1}A_d-I=W_d-I$. Consequently, if the graph is connected, $\lambda_1=1$ is a simple eigenvalue of the weighted adjacency matrix.
%}
%\end{proof}

%Simply we can show that $\Delta^{-1}A_d-I$ is a Laplacian matrix. Consequently for a connected topology we know that $0$ is a simple eigenvalue of this matrix. It is straightforward that $1$ is a simple eigenvalue of the weighted adjacency matrix. Without loosing the generality put the eigenvalues in order and therefore $\lambda_1=1$.

In the rest of this paper we assume that the graph $\mathcal{G}$ describing the communication topology is {\em connected}. By Lemma~\ref{lemma_connected} this implies that its largest eigenvalue is $\lambda_1=1$. We call \emph{unitary eigenvalue mode} (UEM) the mode associated with $\lambda_1=1$.

%\section{The unitary eigenvalue mode}\label{sec:UEM}

The following lemma characterizes the dynamics of the UEM. In particular it shows that the UEM converges asymptotically to a vector whose first entry $z_1(t)$ is equal to a constant value and the second entry $\dot z_1(t)$ is null.

\begin{lemma}\label{lemma1}
Consider a system whose dynamics in the time interval $t\in [t_k,t_{k+1})$, $k=0,1,\ldots,\infty$, follows eq.~\eqref{eq1} with $\lambda_i=1$. Assume $t_{k+1}-t_k>0$ for any $k=0,1,\ldots,\infty$. It holds

\begin{equation}
\lim_{k\rightarrow\infty} z_1(t_k)=\gamma, \quad \gamma \in \rea.
\end{equation}
\end{lemma}

{\em Proof:} To prove this lemma we observe that by eq.~\eqref{eq1} and by definition of matrices $A$ and $B$, it follows that
\begin{equation}\label{equation_f}
\ddot{z_1}(t)+k_d \dot{z_1}(t)+k_p z_1(t)=k_d \dot{z_1}(t_k)+k_p z_1(t_k),
\end{equation}
for $t \in [t_k \ \ t_{k+1}]$. We consider two cases separately.

\subsection*{Case A}

The characteristic polynomial associated with eq.~\eqref{equation_f} has two distinct roots. This corresponds to
$$\sigma = \dfrac{k_d^2}{4}-k_p\ne 0. $$

In such a case the solution of the above ordinary linear differential equation is equal to:
\begin{equation}\label{equation_f_solution}
\begin{array}{lll}
z_1(t) & = & c_1 \dot{z_1}(t_k) e^{s_1 (t-t_k)}-c_2 \dot{z_1}(t_k)e^{s_2 (t-t_k)} \\ & & +z_1(t_k)+\displaystyle \frac{k_d}{k_p}\dot{z_1}(t_k),
\end{array}
\end{equation}
where
$$\begin{array}{c}
s_{1,2}=\dfrac{-k_d}{2}\pm \sqrt{\dfrac{k_d^2}{4}-k_p},\\
 c_1=\dfrac{1}{s_1-s_2}(1+\dfrac{k_d}{k_p}s_2),  \\
 c_2=\dfrac{1}{s_1-s_2}(1+\dfrac{k_d}{k_p}s_1).
\end{array} $$
Now, let $T_k=t_{k+1}-t_k$. From \eqref{equation_f_solution} we can compute $z_1(t_{k+1})$ and $\dot{z_1}(t_{k+1})$ as:
\begin{equation}\label{f_statespace}
\left[\begin{array}{c}
z_1(t_{k+1}) \\
\dot{z_1}(t_{k+1})
\end{array} \right]=M(T_k)\left[\begin{array}{c}
z_1(t_k) \\
\dot{z_1}(t_k)
\end{array}\right]
\end{equation}
where
\begin{equation}
M(T_k)=\left[\begin{array}{cc}
 1&\mu_k  \\
 0 & \beta_k
\end{array} \right],
\end{equation}
\begin{equation}
\mu_k=c_1 e^{s_1T_k}-c_2e^{s_2T_k}+\dfrac{k_d}{k_p},
\end{equation}
and
\begin{equation}\label{def:beta}
\beta_k=c_1s_1 e^{s_1T_k}-c_2s_2e^{s_2T_k}.
\end{equation}
Therefore for all $k>0$ it holds:
\begin{equation*}\label{f_statespace_k0}
\left[\begin{array}{c}
{z_1}(t_{k}) \\
\dot{z_1}(t_{k})
\end{array} \right]=\bar{M}_{k}\left[\begin{array}{c}
{z_1}(0) \\
\dot{z_1}(0)
\end{array}\right]
\end{equation*}
where
\begin{equation}\label{Mbar}
\begin{array}{lll}\bar{M}_{k} & = & M(T_k)M(T_{k-1})\ldots M(T_{0}) \\ & = & \left[\begin{array}{cc}
 1&\sum\limits_{m=0}^{k}\mu_m\prod\limits_{j=0}^{m-1}\beta_j  \\
 0 & \prod\limits_{j=0}^{k}\beta_j
\end{array} \right].
\end{array}
\end{equation}
Since for all $j>0$ it is $\rvert\beta_j\lvert<1$ (see Appendix~A) we get:
$$\lim\limits_{k\to \infty}\prod\limits_{j=0}^{k}\beta_j=0.$$
Therefore, due to the fact that for all $m>0$ the norm of $\mu_{m}$ is bounded by some $\bar\mu<\infty$, we can conclude that the term $\sum\limits_{m=0}^{k}\mu_m\prod\limits_{j=0}^{m-1}\beta_j$, which is obtained multiplying bounded numbers and exponentially decreasing products gets a constant bounded value $\bar\Pi$. Hence $\lim\limits_{k\to\infty}z_1(t_{k})=\lim\limits_{t\to\infty}(z_1(0)+\bar{\Pi}\dot{z}_1(0))$ and $\lim\limits_{k\to\infty}\dot{z_1}(t_{k})=0$
 which in turn implies that there exists $\gamma \in \rea$ such that:
\begin{equation}
\lim\limits_{k\to \infty}z_1(t_{k})=\gamma.
\end{equation}

\subsection*{Case B}

The characteristic polynomial
of \eqref{equation_f} has a single real root $s=-k_d/2$ with multiplicity $2$.

In such a case the solution of eq.~\eqref{equation_f} is:
\begin{equation}\label{equation_f_solution_sigma}
\begin{array}{lll}
z_1(t)& = & d_1 \dot{z_1}(t_k) t e^{s_1 (t-t_k)}-d_2 \dot{z_1}(t_k)e^{s_2 (t-t_k)} \\ & & +z_1(t_k)+\displaystyle \frac{k_d}{k_p}\dot{z_1}(t_k),
\end{array}
\end{equation}
where
$$\begin{array}{c}
 d_1=\left(1+\dfrac{k_d}{k_p}s\right) =0 \\
 d_2=\left(t_k+\dfrac{k_d}{k_p}t_k s+\dfrac{k_d}{k_p}\right)=\dfrac{2}{k_d}.
 \end{array} $$
Therefore it is
 \begin{equation}\label{f_statespace_sigma}
 \left[\begin{array}{c}
 {z_1}(t_{k+1}) \\
 \dot{z_1}(t_{k+1})
 \end{array} \right]=M^\prime(T_k)\left[\begin{array}{c}
 {z_1}(t_k) \\
 \dot{z_1}(t_k)
 \end{array}\right],
 \end{equation}
 where
 $$M^\prime(T_k)=\left[\begin{array}{cc}
  1&\mu^\prime_k  \\
  0 & \beta^\prime_k
 \end{array} \right],$$
 with $\mu^\prime_k=\dfrac{k_d}{k_p}(1-e^{s T_k})$, and $\beta^\prime_k=-e^{s T_k}$. Since
 for any $T_k>0$, it is $\lvert\beta_k\rvert <1$, then, repeating the same reasoning as in Case~A, we
conclude that there exists $\gamma\in \rea$ such that
\begin{equation}
\lim\limits_{k\to \infty}z_1(t_k)=\gamma .
\end{equation} \hfill $\square$

%\section{Stability of the other modes}\label{sec:stability}

We now characterize the conditions on the design parameters $k_p, k_d, \bar \tau$ under which the modes $y_i(t)$, $i=2,\ldots,n$, defined in eq.~\eqref{def:mode} are exponentially stable.

To do this we provide the following lemma, whose proof is inspired by \cite{seuret2012novel}.

\begin{lemma}\label{LMI:theorem}
Consider the generic mode $y_i(t)$ defined in eq.~\eqref{def:mode} whose dynamics follows eq.~\eqref{mode_dynamics}. Matrices $A$, $B$ are defined as in eq.~\eqref{def:AB}, $\tau(t)=t-t_k$, $k=0,1,\ldots,\infty$, and $\lambda_i \in [-1,1)$.

Assume that the difference between any two consecutive sampling times is smaller than a given $\bar{\tau}$, i.e., it holds $t_{k+1}-t_k \leq \bar \tau$ for all $k=0,1,\ldots,\infty$.

If there exist symmetric positive definite matrices $P_i$, $R_i$, $S_i \in \rea ^{2 \times 2}$, a matrix $Q_i=\left[\begin{array}{c}
  Q_{i,1}\\Q_{i,2}
  \end{array} \right] \in\rea ^{4 \times 2}$ and a constant value $\alpha>0$ such that the following inequalities are satisfied:
 \begin{equation}\label{LMI_main_1}
   \left[\begin{array}{cc}
    \Psi_{i,11}(\bar{\tau},\alpha)& \Psi_{i,12}(\bar{\tau},\alpha)\\
    *&\Psi_{i,22}(\bar{\tau},\alpha)
   \end{array} \right]< 0,
  \end{equation}
 \begin{equation}\label{LMI_main_2}
  \left[\begin{array}{ccc}
      \Psi_{i,11}(0,\alpha)& \Psi_{i,12}(0,\alpha)& \bar{\tau}Q_{i,1}\\
      *&\Psi_{i,22}(0,\alpha) &\bar{\tau}Q_{i,2}\\
      *&*&-\bar{\tau}(1-2\alpha\bar{\tau})R_i
     \end{array} \right]< 0
 \end{equation}
 where
 $$\begin{array}{l}
  \begin{array}{l}
   \Psi_{i,11}(\bar{\tau},\alpha)= P_iA+A^TP_i-S_i-Q_{i,1}-Q_{i,1}^T \\ \qquad +
   \bar{\tau}(S_iA+A^TS_i+A^TR_iA+2\alpha S_i)\\ \qquad +2\alpha P_i-2\alpha R_i ,
  \end{array}  \\~\\
  \begin{array}{l}
  \Psi_{i,12}(\bar{\tau},\alpha)= \lambda_i P_iB+S_i+2\alpha R_i+Q_{i,1}-Q_{i,2}^T \\ \qquad +
    \bar{\tau}(-A^TS_i+\lambda_i S_iB+\lambda_i A^TR_iB-2\alpha S_i),
  \end{array}
  \\~\\
  \begin{array}{l}
  \Psi_{i,22}(\bar{\tau},\alpha)=  -S_i-2\alpha R_i +Q_{i,2}+Q_{i,2}^T \\ \qquad
  -\bar{\tau}(\lambda_i B^TS+\lambda_i S_iB-\lambda_i^2 B^TR_iB+2\alpha S_i),
  \end{array}
 \end{array}
 $$
then mode $y_i(t)$ is exponentially stable with decay rate $\alpha$.
\end{lemma}

{\em Proof:} Consider the following functional:
\begin{equation}\label{def_V1}
\begin{array}{lll}
V_i(t,y_i(t),y_i(t_k)) &= & y_i^T(t)P_i y_i(t) \\ & &  +\left(\bar{\tau}-\tau(t)\right) \xi_i^T(t)S_i\xi_i(t) \\
 & & +\left(\bar{\tau}-\tau(t)\right)\int_{t_k}^{t}\dot{y_i}^T(s)R_i\dot{y_i}(s)ds,
\end{array}
\end{equation}
where
\begin{equation}
\xi_i(t)=y_i(t)-y_i(t_k).
\end{equation}
Obviously $\dot{\xi}_i(t)=\dot{y}_i(t)$. {Note that the second and the third term of the functional vanish during the jump due to the fact that $\lim\limits_{t\to t_k}y_i(t)=y_i(t_k)$ which leads to $\lim\limits_{t\to t_k}V(t)\leq V(t_k^-)$. Hence we should look the functional only inside the intervals without being worried about the jumps.}

Derivating eq.~\eqref{def_V1} with respect to time we get:
\begin{equation}\label{vdot}
\begin{array}{l}
\dot{V}_i(t,y_i(t),y_i(t_k))=y_i^T(t)\Big(P_iA+A^TP_i-S_i\\ \qquad +
\big(\bar{\tau}-\tau(t)\big)(S_iA+A^TS_i+A^TR_iA)\Big)y_i(t)\\ \qquad
+2y_i^T(t)\Big(\lambda_iP_iB+S_i+\big(\bar{\tau}-\tau(t)\big)(S_iA \\ \qquad +A^TS_i+A^TR_iA)\Big)y_i(t_k) \\ \qquad + y_i^T(t_k)\Big( -S_i
  -\big(\bar{\tau}-\tau(t)\big)(\lambda_i B^TS_i+\lambda_i S_iB \\ \qquad -\lambda_i^2 B^TR_iB\Big)y_i(t_k)-\int_{t_k}^{t}\dot{y}_i^T(s) R_i \dot{y}_i(s)ds.
%\xi_i^T(t)\big[ 2M_1^TPM_3-M_2SM_2\\
%+\left(\bar{\tau}-\tau(t)\right)\left(2M_3^T S M_2+M_3 RM_3\right)\big]\xi_i(t) \\
\end{array}
\end{equation}
Now consider the following candidate functional:
\begin{equation}\label{Wdot}
\begin{array}{l}
W_i(t,y_i(t),y_i(t_k),\alpha) \\ \qquad =
\dot{V}(t,y_i(t),y_i(t_k))+2\alpha V_i(t,y_i(t),y_i(t_k))\\ \qquad =
y_i^T(t)\Big(P_i A+A^T P_i-S_i+2\alpha P_i \\ \qquad +
\big(\bar{\tau}-\tau(t)\big)(S_iA+A^TS_i+A^TR_iA+2\alpha S_i)\Big)y_i(t)\\
\qquad +2y_i^T(t)\Big(\lambda_iP_iB+S_i+\big(\bar{\tau}-\tau(t)\big)(S_iA \\ \qquad +A^TS_i+A^TR_iA-2\alpha S_i)\Big)y_i(t_k)\\ \qquad + y_i^T(t_k)\Big( -S_i
  -\big(\bar{\tau}-\tau(t)\big)(\lambda_i B^TS_i+\lambda_i S_iB\\ \qquad -\lambda_i^2 B^TR_iB+2\alpha S_i\Big)y_i(t_k)\\ \qquad -(1-2\alpha (\bar{\tau}-\tau(t)))\int_{t_k}^{t}\dot{y}_i^T(s) R_i \dot{y}_i(s)ds.
%\xi_i^T(t)\big[ 2M_1^TPM_3-M_2SM_2\\
%+\left(\bar{\tau}-\tau(t)\right)\left(2M_3^T S M_2+M_3 RM_3\right)\big]\xi_i(t) \\
\end{array}
\end{equation}
To ensure the exponential stability of mode $y_i(t)$ with decay rate $\alpha$ it is sufficient to prove that:
$$W_i(t,y_i(t),y_i(t_k),\alpha)<0.$$

We manipulate the integral term
\begin{equation}
-(1-2\alpha (\bar{\tau}-\tau(t)))\int_{t_k}^{t}\dot{y}_i^T(s) R_i \dot{y}_i(s)ds
\end{equation}
to achieve a bound on that based on a function of $y_i(t)$ and $y_i(t_k)$. To this aim, we rewrite the above term as
the summation of two terms
\begin{equation}\label{int1}
-(1-2\alpha \bar{\tau})\int_{t_k}^{t}\dot{y}_i^T(s) R_i \dot{y}_i(s)ds
\end{equation}
and
\begin{equation}\label{int2}
-2\alpha \tau(t) \int_{t_k}^{t}\dot{y}_i^T(s) R_i \dot{y}_i(s)ds
\end{equation}
and provide an upper bound to each term separately.

To provide an upper bound to \eqref{int1}, we introduce the following inequality for two vectors $\omega_1$ and $\omega_2$ and an arbitrary matrix $\Gamma$ with compatible dimensions:
$$2\omega_1^T\omega_2\le \omega_1^T \Gamma^{-1} \omega_1+\omega_2^T\Gamma \omega_2.$$
Rewriting the above inequality assuming $\omega_1=Q_i^T \left[\begin{array}{c}
 y_i(t)\\
 y_i(t_k)
\end{array}\right]$, $\omega_2=\dot{y_i}(s)$ and $\Gamma=(1-2\alpha\bar{\tau})R_i $, we get:
$$ \begin{array}{c}
2[y_i^T(t)\ \ y_i^T(t_k)]Q_i\dot{y_i}(s)\leq\\
\left[y_i^T(t)\ \ y_i^T(t_k)\right]Q_i \dfrac{R_i^{-1}}{1-2\alpha \bar{\tau}}Q_i^T \left[\begin{array}{c}
 y_i(t)\\  y_i(t_k)
\end{array}\right]\\ +(1-2\alpha \bar{\tau})\dot{y_i}^T(s)R_i\dot{y_i}(s).
\end{array}$$
Integrating it in the interval $\left[t_k, \ t\right]$ in which $\dot{y_i}(t)$ is continuous we obtain:
\begin{equation}\label{inequality}
\begin{array}{c}
-(1-2\alpha \bar{\tau})\int_{t_k}^{t}\dot{y}_i^T(s) R \dot{y}_i(s)ds \leq \\ -2[y_i^T(t)\ \ y_i^T(t_k)]Q_i\xi_i(t)\\
+\tau(t)\left[y_i^T(t)\ \ y_i^T(t_k)\right]Q_i \dfrac{R_i^{-1}}{1-2\alpha \bar{\tau}}Q_i^T \left[\begin{array}{c}
 y_i(t)\\  y_i(t_k)
\end{array}\right].
\end{array}
\end{equation}
%Knowing that $\xi_i(t)-\int_{t_k}^{t}\dot{y_i}(s)ds=0$ leads us to:
%\begin{equation*}
%2[y_i^T(t)\ \ y_i^T(t_k)]Q_i\xi_i(t)=2[y_i^T(t)\ \ y_i^T(t_k)]Q_i \int_{t_k}^{t}\dot{y_i}(s)ds.
%\end{equation*}
%which can be replaced as the left side of the inequality \eqref{label}
%Adding \eqref{inequality} to \eqref{vdot}, the following inequality is achieved for $t\in\left[t_k, \ t_{k+1}\right) $:
%\begin{equation}\label{LMI_with_tau}
%\begin{array}{l}
% \dot{V}_i(t,y_i(t),y_i(t_k))\leq[y_i^T(t) \ \ y_i^T(t_k)] \\
% \Big(\left[\begin{array}{cc}
%       \Psi_{i,11}(\bar{\tau}-\tau(t),0)& \Psi_{i,12}(\bar{\tau}-\tau(t),0)\\
%       *&\Psi_{i,22}(\bar{\tau}-\tau(t),0)
%      \end{array} \right]\\+\tau(t)(Q_iR_i^{-1}Q_i^T-\Phi_i)\Big)  \left[\begin{array}{c}
%           y_i(t)\\  y_i(t_k)
%          \end{array}\right]
%\end{array}.
%\end{equation}

To provide an upper bound to \eqref{int2} we use Jensen integral inequality {\cite{xu2008survey}}:
\begin{equation}\label{jensen}
\begin{array}{c}
-2\alpha\tau(t)\int\limits_{t_k}^{t}\dot{y}_i^T(s)R_i\dot{y}_i(s)ds \leq \\ -2\alpha\int\limits_{t_k}^{t}\dot{y}_i^T(s)ds R_i\int\limits_{t_k}^{t}\dot{y}_i(s)ds  \\=-2\alpha(y_i(t)-y_i(t_k))^TR_i (y_i(t)-y_i(t_k))
\end{array}
\end{equation}

Introducing inequalities \eqref{inequality} and \eqref{jensen} in \eqref{Wdot}, the following inequality is achieved for $t\in\left[t_k, \ t_{k+1}\right) $:
\begin{equation}\label{LMI_with_tau_W}
\begin{array}{l}
 {W}_i(t,y_i(t),y_i(t_k))\leq[y_i^T(t) \ \ y_i^T(t_k)] \\
 \Big(\left[\begin{array}{cc}
       \Psi_{i,11}(\bar{\tau}-\tau(t),\alpha)& \Psi_{i,12}(\bar{\tau}-\tau(t),\alpha)\\
       *&\Psi_{i,22}(\bar{\tau}-\tau(t),\alpha)
      \end{array} \right]\\+\dfrac{\tau(t)}{1-2\alpha\bar{\tau}}Q_iR_i^{-1}Q_i^T\Big)  \left[\begin{array}{c}
           y_i(t)\\  y_i(t_k)
          \end{array}\right].
\end{array}
\end{equation}
The above inequality corresponds to an LMI that is linear with respect to $\tau(t)$. Therefore, according to \cite{scherer2000linear}, in order to be
sure that it holds for all $\tau(t)\in [0, \ \bar{\tau} ]$ we only need to check it at the boundary of the interval, namely for
$\tau(t)=0$ and $\tau(t)=\bar{\tau}$.

Now, if we particularize eq.~\eqref{LMI_with_tau_W} with $\tau(t)=0$ this obviously leads to the LMI in eq.~\eqref{LMI_main_1}.

To complete the proof we need to show that particularizing eq.~\eqref{LMI_with_tau_W} with $\tau(t)=\bar \tau$ we get the LMI in eq.~\eqref{LMI_main_2}. But this follows from the fact that
\begin{equation}\label{Schur_2}
\left[\begin{array}{cc}
       \Psi_{i,11}(0,\alpha)& \Psi_{i,12}(0,\alpha)\\
       *&\Psi_{i,22}(0,\alpha)
      \end{array} \right]\\+\dfrac{\bar \tau}{1-2\alpha\bar{\tau}}Q_iR_i^{-1}Q_i^T
\end{equation}
is the Schur complement of matrix $-\bar \tau (1-2 \alpha \bar \tau)R_i$ in eq.~\eqref{LMI_main_2}. Thus, if the LMI in eq.~\eqref{LMI_main_2} is definite negative, also it is matrix in eq.~\eqref{Schur_2}. \hfill $\square$

\subsection{Consensus among agents}

We now prove the main result of the paper, namely the consensus of agents to a common position.

\begin{theorem}
Consider a MAS evolving according to equation \eqref{eq_2_c} where $\bar \tau$ is such that $0<t_{k+1}-t_k<\bar \tau<\infty$. Let $\lambda_i$, $i=2,\ldots,n$ be the eigenvalues of the weighted adjacency matrix associated with the undirected connected graph $\mathcal{G}$ modeling the communication topology. If there exists a positive constant $\alpha$ such that the LMIs defined in eq.~\eqref{LMI_main_1} and \eqref{LMI_main_2} are satisfied for all $\lambda_i$, $i=2,\ldots,n$, then there exists a $\gamma \in \rea$ such that $x(t)$ exponentially converges to $\gamma \vec{1}$ and $v(t)$ exponentially converges to $\vec{0}$. Moreover, the rate of convergence is greater than or equal to $\alpha$.
\end{theorem}

{\em Proof:} By Lemma~\ref{LMI:theorem}, if the LMIs in eq.~\eqref{LMI_main_1} and \eqref{LMI_main_2} hold, all modes except the UEM are stable, i.e., $\lim\limits_{t\to \infty}y_i(t)=0$ and thus $ \lim\limits_{t\to \infty}z_i(t)=0$ for $i=2,\ldots,n$ with rate of convergence of at least $\alpha$. Furthermore, by Lemma~\ref{lemma1}, there exists a positive constant $\gamma \in \rea$ such that $\lim\limits_{t\to \infty}z_1(t)=\gamma$.

Now, the first column of $T$ is the eigenvector corresponding to the unitary eigenvalue of $W_d$, therefore it is equal to $\vec 1=[1\ \ 1\ \ \ldots,\ \ 1 ]^T$. Thus, being $x(t)=T [z_1(t) \ \ 0 \ \ \ldots \ \ 0]^T$, it is trivial to show that when $t \to \infty$ it is $x_i(t)=x_j(t)$, for all $i,j=1,\ldots, n$.
The same calculations can be repeated for the velocities, thus proving that for $t \to \infty$, it is $v_i(t)=v_j(t)$, $i,j=1,\ldots, n $. \hfill $\square$

\section{Simulation results}\label{sim_res}

In this section we present the results of some numerical simulation that shows the effectiveness of the consensus
protocol in eq.~\eqref{main}.  To this aim we consider a system with $6$ agents and adjacency matrix:
$$A_d=\left[\begin{array}{cccccc}
0 & 1 &0  &1  &0  &0  \\
     1 & 0 & 0 & 1 & 0 & 0\\
     0 & 0 & 0 & 1 & 0 & 1\\
     1 & 1 & 1 & 0 & 1 & 0\\
     0 & 0 & 0 & 1 & 0 & 1\\
     0 & 0 & 1 & 0 & 1 & 0\end{array}\right]. $$

Assume $k_p=1$, $k_d=2$ and $\bar \tau=1$. Using the above LMIs with $\alpha=0.38$ we can prove that the system reaches consensus to a fixed point.

\begin{figure}
\centering
\includegraphics[width=1\linewidth]{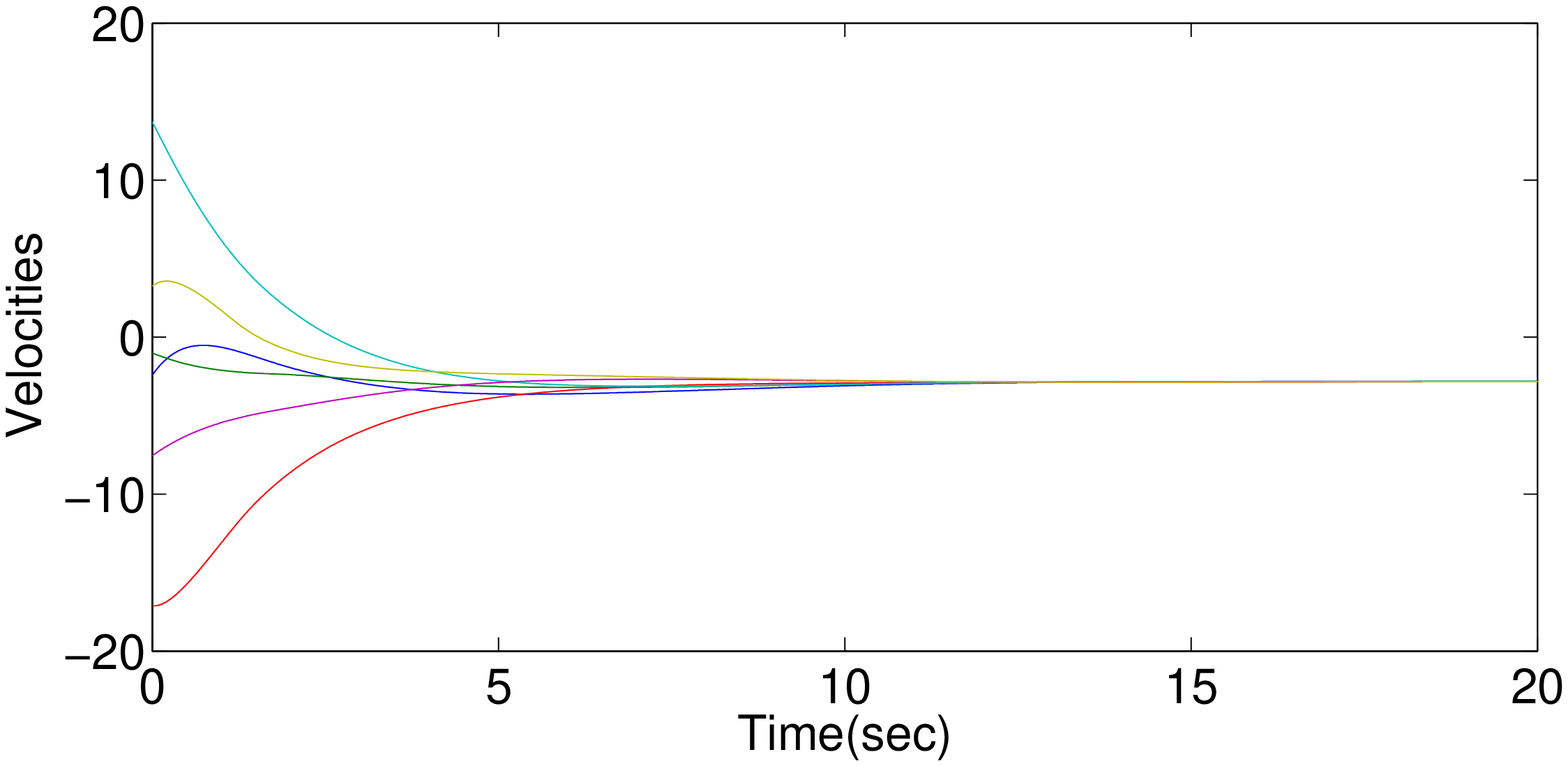}
\includegraphics[width=1\linewidth]{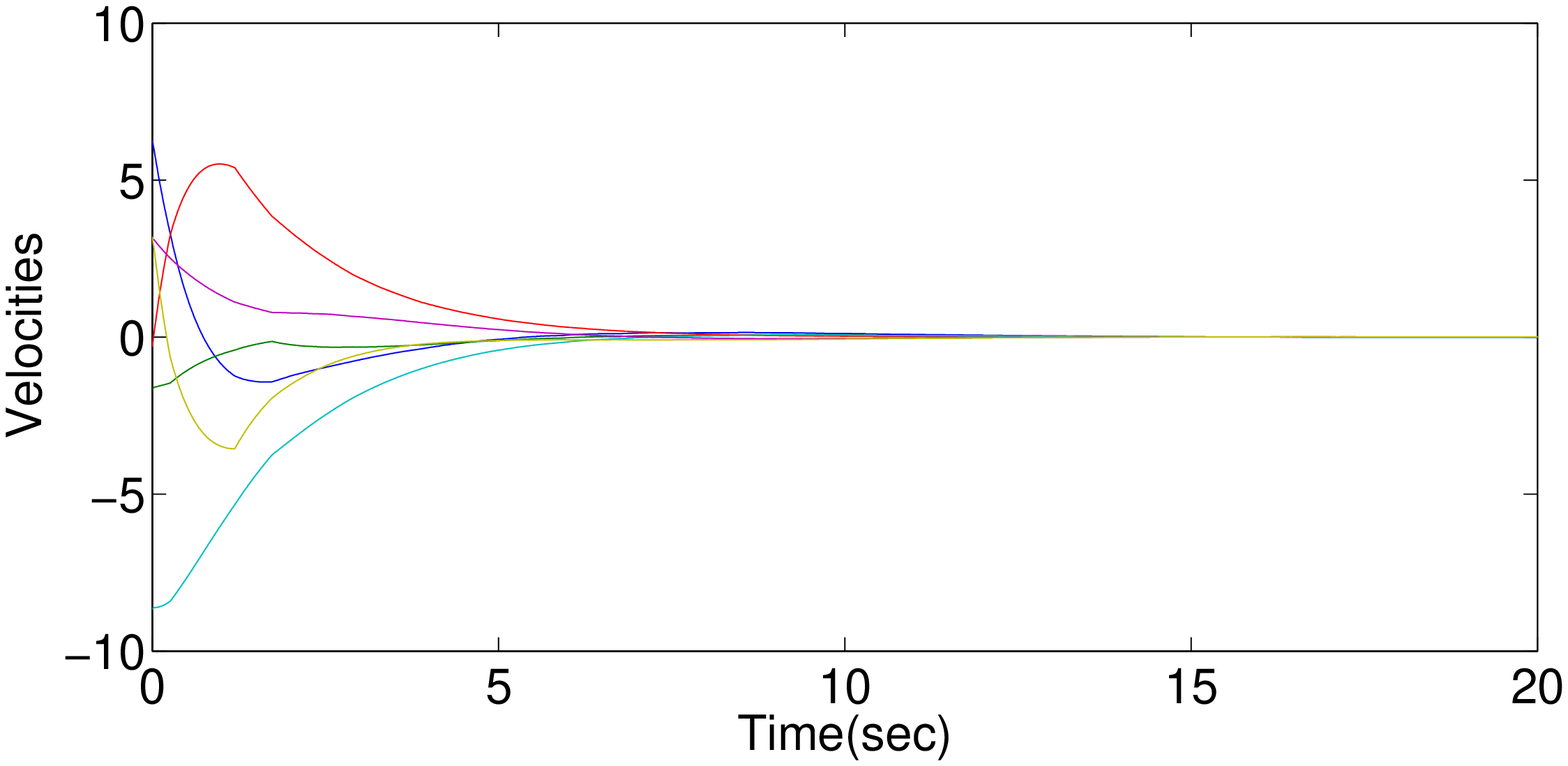}
\caption{Positions and velocities when the proposed protocol is implemented. }
\label{fig:postions}
\end{figure}

\begin{figure}
\centering
\includegraphics[width=1\linewidth]{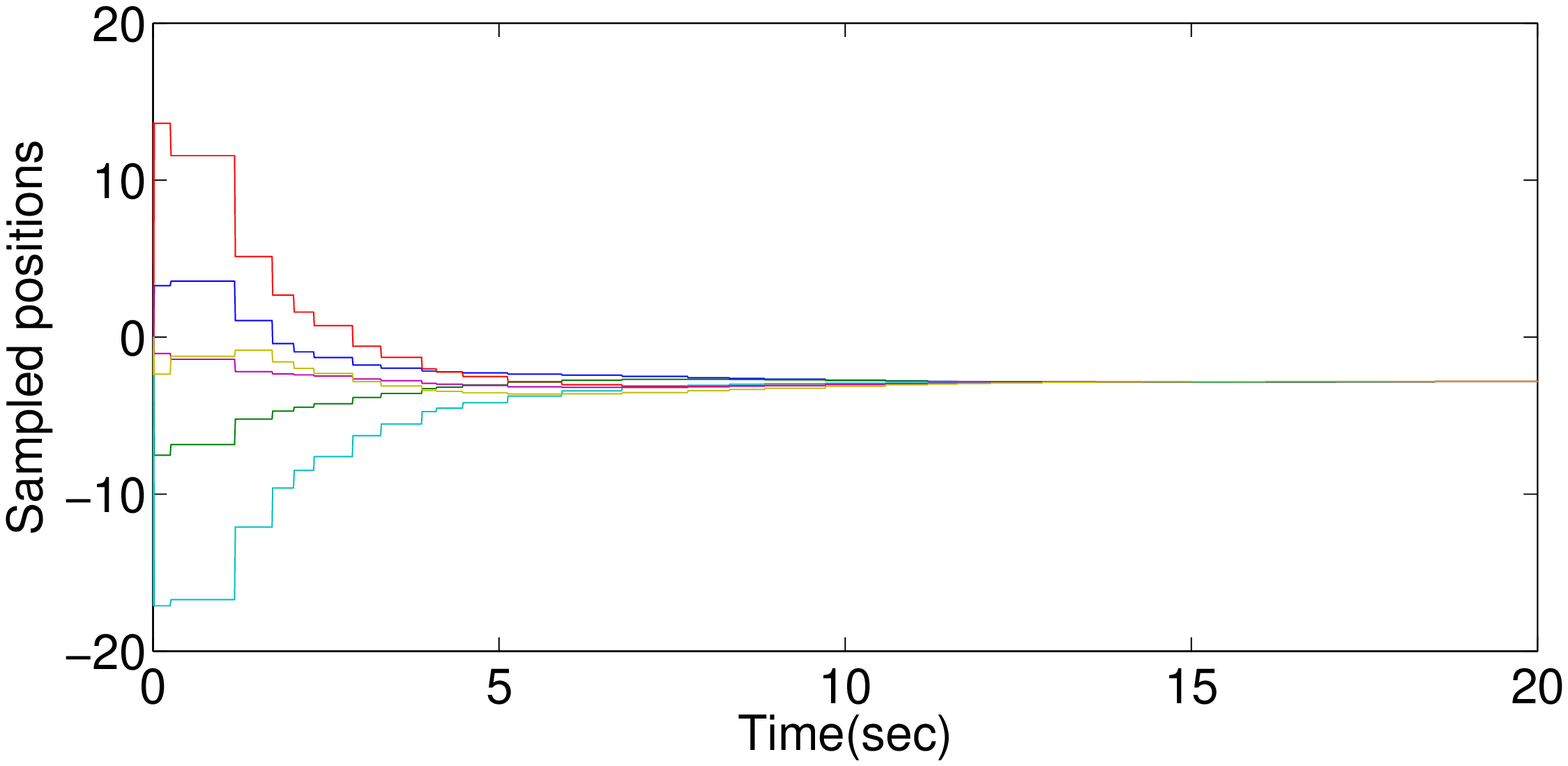}
\includegraphics[width=1\linewidth]{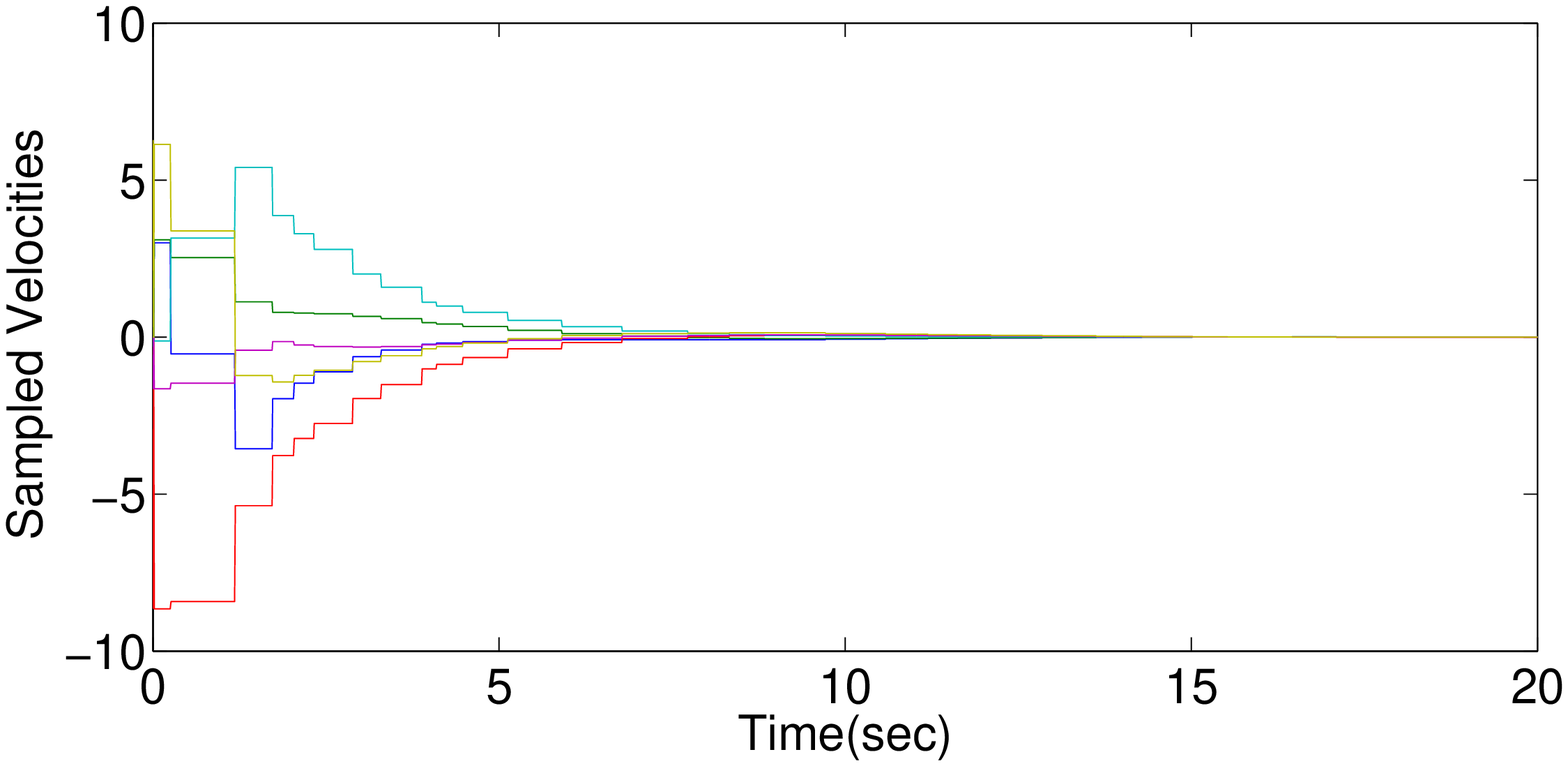}
\caption{Aperiodic sampled positions and velocities when the proposed protocol is implemented.  }
\label{fig:sampledvel}
\end{figure}

Fig.~\ref{fig:postions} shows the evolution of positions and velocities when the proposed algorithm is implemented, while Fig.~\ref{fig:sampledvel} shows the sampled positions and velocities aperiodically transmitted to neighbors by each agent.

We conclude this section presenting the results of another numerical simulation carried out under the assumption that only sampled positions are transmitted to neighbors, i.e., the second term is removed in eq.~\eqref{eq_2_c} that is equivalent to redefine $B$ as $ B^\prime=[
 0 \ 0;  k_p \ 0].$

\begin{figure}
\centering
\includegraphics[width=1\linewidth]{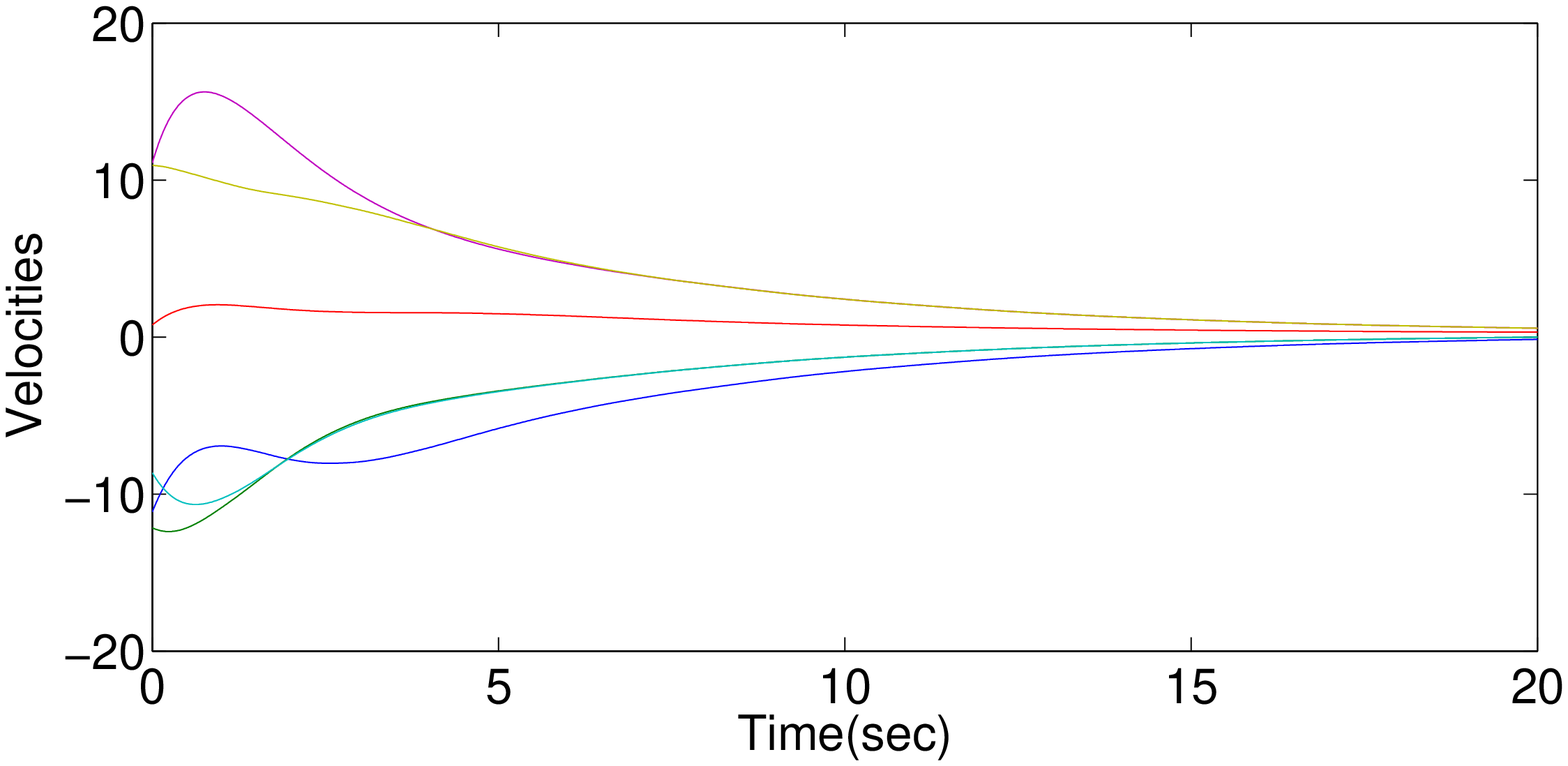}
\includegraphics[width=1\linewidth]{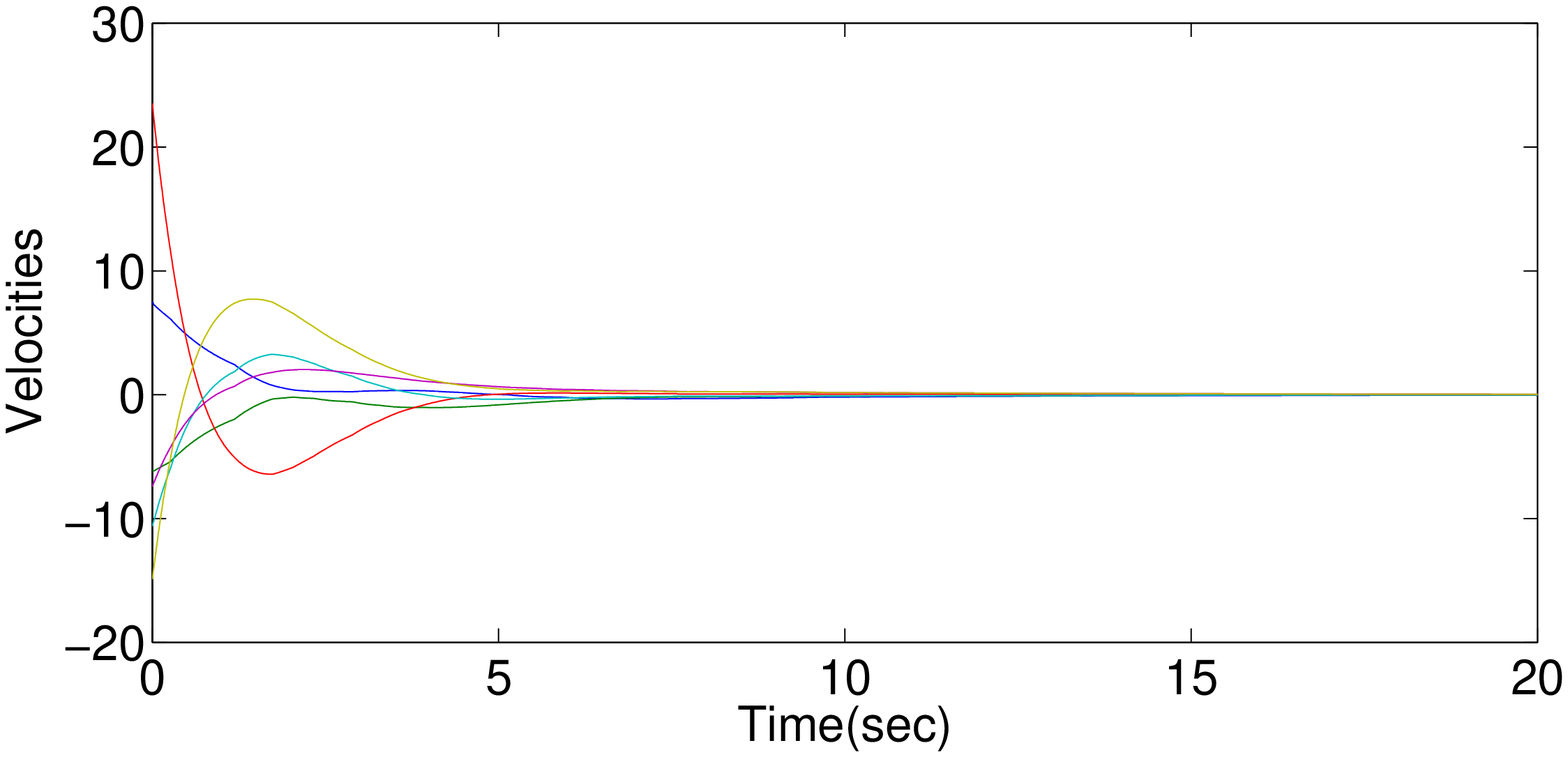}
\caption{Positions and velocities when the proposed protocol is modified in order to only consider sampled positions. }
\label{fig:postions_pos_sampled}
\end{figure}

It can be proved that in such a case the consensus to a fixed point is still reached, but with decay rate bounded by $0.21$ that is almost the half of the previous case. Such a conclusion can also be drawn by looking at Fig.~\ref{fig:postions_pos_sampled}.

\section{Conclusions and future work}\label{conclusions}
The contribution of this paper consists in a PD-like consensus algorithm for a second-order multi-agent system where, at non-periodic sampling times, agents transmit to their neighbors information about their position and velocity, while each agent has a perfect knowledge of its own state at any time instant. Conditions have been given to prove consensus to a common fixed point, based on LMIs verification. Moreover, we also show how it is possible to evaluate an upper bound on the decay rate of exponential convergence of stable modes.

Three are the main directions of our future research in this framework.

--- First, we plan to provide sufficient conditions for consensus on graphs whose spectrum is not known, but only a measure of the connectivity is given. %To this aim we plan to rewrite the LMI coming from eq.~\eqref{LMI_with_tau_W} as a linear function of both $\tau(t)$ and $\lambda_i$. This enables us to use the results in \cite{scherer2000linear} simultaneously for $\tau(t)$ and $\lambda_i$, thus not requiring the exact knowledge of all $\lambda_i$'s, but only an upper bound on them. Moreover, in such a case, the number of LMIs that should be verified is not equal to $2(n-1)$, where $n$ is the number of agents, but depends on the number of vertices of the polynomial defining the admissible range of parameters that is necessary to make the LMI linear with respect to $\tau(t)$ and $\lambda_i$, that does not depend on $n$.

--- Second, we want to compute analytically an upper bound on the value of the second largest eigenvalue of the weighted adjacency matrix that guarantees consensus, as a function of the other design parameters. %To this aim we plan to repeat the calculations in the proof of Lemma~\ref{lemma1} considering the dynamics of the other modes.

--- Third, we want to also study the case where agents do not have a perfect knowledge of their own state.

--- Finally, we plan to relax the assumption that all communications among agents occur simultaneously.

\bibliographystyle{plain}
\bibliography{biblio}

\section*{Appendix A}

We now prove that $\lvert\beta_k\rvert<1$ where $\beta_k$ is defined as in eq.~\eqref{def:beta}.

Let $$\displaystyle s_1=\frac{-k_d}{2}+\sqrt{\sigma}, \ \ s_2=\frac{-k_d}{2}+\sqrt{\sigma}, \ \ \sigma=\frac{k_d^2}{4}-k_p.$$
We consider separately the case of $\sigma>0$ and $\sigma<0$.

\subsection*{Case~1: $\sigma>0$}
 In this case it is trivial to show that $s_1, s_2 \in \mathbb{R}$ and $s_2<s_1<0$.
 Furthermore, we have $e^{s_2T_k}<e^{s_1T_k}$ and $c_1s_1-c_2s_2=1$. We can also show that:
 \begin{equation*}
 \begin{array}{ll}
    c_1s_1 & =  \dfrac{1}{s_1-s_2}(s_1+\dfrac{k_d}{k_p} s_1 s_2) \\
   & = \dfrac{1}{2\sqrt{\sigma}}(\dfrac{k_d}{2}+\sqrt{\sigma})>0
  \end{array}
 \end{equation*}
 and
 \begin{equation*}
  \begin{array}{ll}
     c_2s_2 & =  \dfrac{1}{s_1-s_2}(s_2+\dfrac{k_d}{k_p} s_1 s_2) \\
    & = \dfrac{1}{2\sqrt{\sigma}}(\dfrac{k_d}{2}-\sqrt{\sigma})>0.
   \end{array}
  \end{equation*}
Let $\omega=\sqrt{\sigma}$ and $\nu=k_d/2=\sqrt{\omega^2+k_p}$. We get:
 \begin{equation}\label{beta_nu}
 \beta_k=\dfrac{(\nu+\omega)e^{\omega T_k}-(\nu-\omega)e^{-\omega T_k}}{2\omega e^{\nu T_k}}.
 \end{equation}
Moreover, since $\sigma>0$, it is $\omega \in (0, \ \infty)$ and therefore $\nu\in (\sqrt{k_p}, \ \infty)$. For any $k_p>0$ we obtain:
\begin{eqnarray*}
\lim\limits_{\omega \to 0}&\beta_k&=\frac{1+\sqrt{k_p }T_k}{e^{\sqrt{k_p }T_k}},\\
\lim\limits_{\omega \to \infty}&\beta_k&=1.
\end{eqnarray*}

Hence due to the continuity in \eqref{beta_nu}, for any value of $k_p$ and $k_d$ such that $\sigma>0$, knowing that $T_k>0$, we achieve
$$\beta_k \in \left( \frac{1+\sqrt{k_p}}{e^{\sqrt{k_p}}},  \ 1\right)$$
thus proving the statement.

 \subsection*{Case~2: $\sigma<0$}

In such a case $s_1$ and $s_2$ are complex conjugate numbers and
 \begin{equation*}
 \begin{array}{ll}
 \beta_k= & (c_1s_1-c_2s_2)e^{-T_k k_d/2 }\cos(\sqrt{\sigma }T_k)+ \\ & j(c_1s_1+c_2s_2)e^{-T_k k_d/2 }\sin(\sqrt{\sigma} T_k).
 \end{array}
 \end{equation*}
Being $c_1s_1+c_2s_2=0$ and $c_1s_1-c_2s_2=1$ the second term vanishes and we get:
 \begin{equation}
  \beta_k=e^{-T_k k_d/2}\cos(\sqrt{\sigma})<1
  \end{equation}
thus proving the statement.

\end{document}